# Graph neural network framework for energy mapping of hybrid monte-carlo molecular dynamics simulations of Medium Entropy Alloys

Mashaekh Tausif Ehsan[1], Saifuddin Zafar[1], Apurba Sarker[1], Sourav Das Suvro[1], Mohammad Nasim Hasan[1*]

[1]Department of Mechanical Engineering, Bangladesh University of Engineering and Technology (BUET), Dhaka-1000, Bangladesh.

[*]Corresponding author *Email address: nasim@me.buet.ac.bd*

Email: mashaekh.tausif@gmail.com, saifuddinzafar16@gmail.com, apurba.sarker.as@gmail.com, sdsuvro99@gmail.com, nasim@me.buet.ac.bd

[*]**Corresponding Author:**

Prof. Mohammad Nasim Hasan

Department of Mechanical Engineering

Bangladesh University of Engineering and Technology (BUET)

Dhaka-1000, Bangladesh

E-mail: nasim@me.buet.ac.bd

Telephone: +8801921506445

# Graph neural network framework for energy mapping of hybrid monte-carlo molecular dynamics simulations of Medium Entropy Alloys

**Abstract:** Machine learning (ML) methods have drawn significant interest in material design and discovery. Graph neural networks (GNNs), in particular, have demonstrated strong potential for predicting material properties. The present study proposes a graph-based representation for modeling medium-entropy alloys (MEAs). Hybrid Monte-Carlo molecular dynamics (MC/MD) simulations are employed to achieve thermally stable structures across various annealing temperatures in an MEA. These simulations generate dump files and potential energy labels, which are used to construct graph representations of the atomic configurations. Edges are created between each atom and its 12 nearest neighbors without incorporating explicit edge features. These graphs then serve as input for a Graph Convolutional Neural Network (GCNN) based ML model to predict the system's potential energy. The GCNN architecture effectively captures the local environment and chemical ordering within the MEA structure. The GCNN-based ML model demonstrates strong performance in predicting potential energy at different steps, showing satisfactory results on both the training data and unseen configurations. Our approach presents a graph-based modeling framework for MEAs and high-entropy alloys (HEAs), which effectively captures the local chemical order (LCO) within the alloy structure. This allows us to predict key material properties influenced by LCO in both MEAs and HEAs, providing deeper insights into how atomic-scale arrangements affect the properties of these alloys.

# 1. Introduction

Equiatomic multi-principal element alloys have emerged as a topic of considerable interest due to their unique combination of mechanical properties, damage tolerance, and radiation resistance [1–4]. When these alloys consist of three or four principal elements, they are classified as medium entropy alloys (MEA). A particularly well-studied MEA system is the NiCoCr alloy, which is derived from the broader FeCoCrMnNi (Cantor) alloy system [5]. The NiCoCr alloy is characterized by a single-phase face-centered cubic structure, with short-range chemical ordering that varies depending on annealing temperature and lattice distortions caused by the presence of different atomic species [6]. Numerous studies highlight the outstanding strength, ductility, and fracture toughness of NiCoCr MEAs, especially at low temperatures, positioning them as highly suitable for demanding applications where superior damage tolerance is required [7–9].

The high configurational entropy in HEAs and MEAs often leads to the formation of single-phase solid solutions. While it is initially believed that the atomic distribution in these alloys is random, recent experimental studies have demonstrated the presence of LCO [10,11]. According to Gibbs free energy principles, at lower temperatures, enthalpy plays a dominant role, causing variations in the affinity between different element pairs, which leads to element segregation and the formation of atomic clusters. The influence of LCO on the properties of HEAs has become a critical research focus in recent years. A recent experimental study identified the presence of LCO in NiCoCr MEAs using energy-filtered transmission electron microscopy. The microscopy images revealed a pronounced avoidance of Cr-Cr pairs at nearest-neighbor distances. Follow-up bulk tensile tests further demonstrated the strengthening effect of LCO, showing a 25\% increase in yield strength in NiCoCr MEAs because of this atomic arrangement [12]. The small scale of LCO presents significant challenges for direct observation through experimental techniques. To address this issue, hybrid MC/MD simulation serves as an effective tool for modeling ordered structures. In this approach, atoms are randomly selected and swapped, ultimately yielding a thermally stable and ordered structure [13–20].

The application of ML in materials science has revolutionized the way researchers approach the discovery of and design of new materials. One of the primary objectives of ML in material science is the prediction of material properties. ML models, trained on large dataset of material properties, have demonstrated the ability to predict complex behavior and properties, reducing the need for extensive and costly experiments [21–23]. More specifically, Neural Networks have been successfully employed to develop interatomic potentials, including the Neural Network Potential (NNP) and the Deep Potential, which enable efficient MD simulations without the necessity of quantum mechanical calculations. [24,25]. The application of Convolutional Neural Networks (CNNs) in predicting material properties has garnered significant attention in recent years. Ahmed et al. have investigated the potential of CNNs for predicting diffusion barriers for materials such as graphene and boron nitride [26]. Cao et al. suggested a hybrid approach that combines material descriptors with CNNs to anticipate crystal material properties, underscoring the significance of feature engineering in improving model performance [27]. Cho and Lin have also demonstrated the use of 3D CNNs to predict the adsorption properties of nanoporous materials, which is a recent advancement. This study demonstrates the potential of CNNs to process and analyze three-dimensional structural data, a critical component of comprehending intricate material behaviors [28].

GNNs have emerged as a particularly effective ML architecture in material science due to their ability to convert atomic structures into graphs. In these models, atoms are represented as nodes and the bonds between them are represented as edges, allowing for the prediction of material properties based on atomic configurations [29–31]. Moreover, Chen et al. introduced the Materials Graph Network (MEGNet), which integrates various atomic and bond attributes along with state attributes such as temperature and pressure [32]. Their findings indicated that MEGNet outperformed previous graph-based models in predicting a wide range of material properties for both molecules and crystals [33]. Additionally, Juan et al. proposed constrained learning techniques to train stable GNNs, which is particularly important in materials science where data can be sparse or noisy [34]. One of the notable advancements of the field is the development of Graph Neural Network Force Field (GNNFF) that can effectively model interatomic forces thereby, enhancing the accuracy of molecular dynamics simulations [35].

In this work, we first perform hybrid MC/MD simulations on a random equiatomic NiCoCr alloy across different annealing temperatures to generate the required dataset. Then, we extract dump files and potential energy values at various simulation steps. These dump files are then converted into edge lists and node feature matrices, enabling the construction of graph representations of the atomic structure. The graph data is subsequently fed into a GCNN-based architecture, which is trained to predict potential energy profiles.

## 2. Graph Convolutional Neural Network

Convolution is a notion that is first established for grid-like data, such as images. GCNNs expand this concept to graph-structured data. Nodes in a graph represent features, while edges in a graph represent the relationships between those features [36]. In complex systems like social networks, chemical structures, or atomic configurations in materials science, where interactions between elements are non-Euclidean and cannot be characterized by fixed coordinates, GCNNs are especially effective.

Consider a graph $G = (V, E)$, where $V$ is a set of nodes and $E$ is a set of edges. Let $A \in R^{N \times N}$ represent the adjacent matrix of the graph, where $N$ is the number of nodes, and $A_{ij} = 1$ if there is an edge between nodes $i$ and $j$, and 0 otherwise. The node features are represented by a matrix $X \in R^{N \times F}$, where $F$ is the number of features per node. In addition, the degree matrix $D \in R^{N \times N}$ is a diagonal matrix where $D_{ii}$ equals the degree of the node $i$, i.e., the number of edges connected to the node $i$.

In traditional CNNs, the convolution operation aggregates local information from neighboring pixels using a filter. GCNNs generalize this operation to graphs by aggregating information from neighboring nodes. The key idea is to perform a "message passing" operation where each node aggregates information from its neighbors to update its feature representation. This operation is typically defined as:

$$H^{(l+1)} = \sigma\left(\widetilde{D}^{-\frac{1}{2}} \widetilde{A} \widetilde{D}^{-\frac{1}{2}} H^{(l)} W^{(l)}\right) \qquad (1)$$

where $H^{(l)}$ is the node feature matrix at the layer $l$, $W^{(l)}$ is a learnable weight matrix for layer $l$, ($\sigma$ is a non-linear activation function, and $\widetilde{A} = A + I$ is the adjacency matrix with added self-loops where $I$ is the identity matrix, which ensures that each node also considers its own feature information. $\widetilde{D}$ is the degree matrix corresponding to $\widetilde{A}$, used to normalize the adjacency matrix, which mitigates the impact of nodes with large degrees dominating the aggregation. The normalizing term $(\widetilde{D}^{-\frac{1}{2}} \widetilde{A} \widetilde{D}^{-\frac{1}{2}})$ is introduced to ensure that the feature representations are appropriately scaled, addressing the problem of varying node degrees across the graph.

**Fig. 1** depicts the flowchart of the architecture for the GCNN-based ML model employed in this study. In this model, we implement the ReLU activation function and apply dropout regularization with a probability set at 0.5. The model is trained using the mean squared error (MSE) loss function and the adaptive moment optimization (ADAM) optimizer [37]. The entire workflow is implemented using PyTorch [38] and PyTorch Geometric [39].

## 3. Methodology

### 3.1 Interatomic potential and simulation cell

The embedded atom method (EAM) potential, developed by [40], is utilized to model atomic interactions. The accuracy and reliability of this potential have been extensively validated in previous studies [41–44]. The EAM potential is expressed as:

$$E_i = F_\alpha \left( \sum_{j \neq i} \rho_\beta(r_{ij}) \right) + \frac{1}{2} \sum_{j \neq i} \phi_{\alpha\beta}(r_{ij}) \tag{2}$$

In this context, the total energy of the atom $i$ is denoted as $E_i$. The term $r_{ij}$ represents the distance between atoms $i$ and $j$, while $\varphi_{\alpha\beta}$ refers to a pairwise potential function. The term $\rho_\beta$ is a functional that depends on the atomic types of both atoms $i$ and $j$, allowing for different elements to contribute variably to the total electron density at a given atomic site based on the identity of the element present at that site. Additionally, $F_\alpha$ serves as an embedding function that quantifies the energy needed to position an atom $i$ within the electron cloud. Here, α and β indicate the types of elements corresponding to atoms $i$ and $j$, respectively.

The random equiatomic NiCoCr structure is created using Atomsk software [45], with a face-centered cubic (FCC) lattice structure and dimensions of 53.52 Å x 53.52 Å x 53.52 Å, containing a total of 13,500 atoms. This structure is called a random solid solution (RSS) system.

### 3.2 Hybrid MC/MD simulation

Hybrid MC/MD simulations are carried out using Large-scale Atomic/Molecular Massively Parallel Simulator (LAMMPS) [46] to achieve energy-minimized LCO configurations. First, the energy of the RSS system is minimized using the conjugate gradient (CG) algorithm, followed by thermal equilibration using the isothermal-isobaric (NPT) ensemble at the annealing temperature $T_a$ and zero pressure for 75 ps. Afterward, the hybrid MC/MD simulation is initiated where two different types of atoms are randomly selected and a swap attempt is made as shown in **Fig. 2(b)**. During each MC step, 10 swap attempts are made to gradually minimize the system's potential energy. A trial swap is accepted if the system energy $E_{i+1}$ after the $i+1$-th swap attempt is lower than the system energy $E_i$ from the previously accepted swap. Otherwise, the acceptance probability is given by [47]

$$P = e^{\frac{E_{i+1}-E_i}{k_B T_a}} \tag{3}$$

where $k_B$ is Boltzmann constant. If the uniformly generated random number within the range (0,1) is less than or equal to $P$, the swap is accepted; otherwise, it is rejected. After attempting 10 swaps on each MC step, the system is relaxed under the NPT ensemble for 2 fs at $T_a$ and zero pressure. A total of 60,000 MC steps are performed for each $T_a$, ranging from 350K to 1150K. The timestep for the simulations is 0.001 ps, and periodic boundary conditions are applied in all three dimensions of the simulation cell. The results of the simulations are visualized using the Open Visualization Tool (OVITO) [48].

### 3.3 Warren-Cowley parameter

The Warren-Cowley (WC) parameter [49] quantifies the degree of LCO in alloys. It measures how likely atoms of different types are to be near each other compared to a random distribution. Mathematically, it's expressed as:

$$\alpha_{ij}^m = 1 - \frac{P_{ij}^m}{c_j} \tag{4}$$

In this context, $m$ denotes the $m$-th nearest-neighbor shell, where $P_{ij}^m$ denotes the average likelihood of finding a $j$-type atom near an $i$-type atom within this $m$-th shell. The term $c_j$ refers to the overall concentration of $j$-type atoms in the system. A negative $\alpha_{ij}^m$ indicates that $j$-type atoms tend to cluster around $i$-type atoms in the $m$-th shell, whereas a positive value suggests a repulsive interaction between the two atom types.

In our study, we focus on the first nearest-neighbor shell ($m$=1). In our NiCoCr MEA, there are three elements, resulting in six different element pairs and, consequently, six distinct WC parameters. An OVITO-Python modifier developed by Sheriff et al. [50] is used to calculate the WC parameters.

## 3.4 Graph construction from dump files

For our GCNN-based ML model, we need to input a graph structure that represents an atomic configuration of the system. To achieve this, we extract dump files and potential energy labels from hybrid MC/MD simulations at intervals of 200 MC steps. Each dump file provides a detailed snapshot of the atomic configuration of the system from which we construct a graph-based representation. In this graph, each atom is considered a node and the bonds between an atom and its 12 nearest neighbor atoms are represented as edges. As there are 13,500 atoms in the system, each graph contains 13,500 nodes. From each dump file, we extract the atom type and absolute velocity of all atoms and create a node feature matrix. Velocity is an important feature which is previously used in GNN to predict the potential energy of a system [51]. To construct the graph, we create an edge list from a dump file where connections are made between each atom (node) and its 12 nearest neighbors, capturing the LCO within the system. Additionally, we include self-loops to ensure that all the information of the nodes is preserved during the message-passing process. Once the node feature matrices and edge lists are generated from each dump file, we convert them into Pytorch Geometric Data objects, a format suitable for use in our GCNN-based architecture. To efficiently manage the training process, we use Pytorch DataLoader to handle batching, shuffling, and loading graph data. We use batch sizes of one. Additionally, we apply shuffling to randomize the order of the graph during the training to prevent the model from overfitting to the sequence of data, thus improving its performance in unseen configurations.

## 3.5 Case studies

We evaluate the performance of our GCNN-based ML model in three different case studies. In the first study, separate models are trained individually on datasets of different annealing temperatures to assess their performance to predict the potential energy at each annealing temperature separately. In the second study, we use a single model trained on a combined dataset of different annealing temperatures to examine how well the model can generalize and predict the potential energy in this range. In the third study, we focus on the model's performance with unseen configurations by training it using some annealing temperatures. The model is then tested on its ability to predict the potential energy across all annealing temperatures. In all cases, the number of epochs is 800, and the learning rate is 0.0001. All the studies are carried out on a GPU-equipped workstation featuring a ZOTAC GAMING GeForce RTX 3060 Ti Twin Edge LHR 8GB with a 6-core/12-thread processor.

### 3.5.1 Model training on specific annealing temperatures

In the first case study, different models are trained individually on datasets corresponding to annealing temperatures of 450K, 650K, and 950K to evaluate the model's ability to predict the potential energy for each temperature independently. As illustrated in **Fig. 1**, The model architecture consists of multiple Graph Convolutional (GCNConv) layers, where each GCNConv layer is followed by a ReLU activation and a dropout layer, forming a GCNConv-ReLU-Dropout block. In the first block, the GCNConv layer has input channels equal to the number of atoms in the system (which is 13500 in our work) and an output channel of 100. After this, there are $n$ GCNConv-ReLU-Dropout blocks, where each GCNConv layer has both input and output channels of 100. In our first case study, $n = 1$. After this, a final GCNConv layer is introduced with a single output channel, which reduces the dimensionality and focuses on the specific task of energy prediction. Subsequently, the output from this GCNConv layer is passed through a fully connected (FC) layer, which serves as the final regression layer to predict the potential energy of the system. When training the model, we store the parameters that achieved the lowest MSE loss, and these optimal parameters are used for the final prediction.

### 3.5.2 Model training on set a of different annealing temperatures

In the second case study, we use a single model to train on a combined dataset consisting of annealing temperatures of 450K, 650K, and 850K. The model architecture starts with the GCNConv-ReLU-Dropout block where the GCNConv layer has an input channel of 13500 and an output channel of 300. After this, there are n number of such blocks and for our second and third case study, $n=3$ and each of the GCNconv layers has an input and output channel of 300. Following these three GCNConv-ReLU-Dropout blocks, a final GCNConv layer with a single output channel is applied. This is followed by an FC layer to output the predicted potential energy. When training the model, we store the parameters of the model that achieved the lowest MSE loss, and these optimal parameters are used for the final predictions.

### 3.5.3 Training of model and testing on unseen configurations

In the third case study, we focus on the model's performance with unseen configurations by training it using data from 350K, 650K, 850K, and 1150K annealing temperatures and the model is then tested to predict the potential energy across all annealing temperatures. We use the same model architecture as in the second case study. During training, we store the parameters that resulted in the lowest MSE loss, and these optimal parameters are then used to predict the potential energies for both seen and unseen configurations across all annealing temperatures.

### 3.6 Evaluation metrics

In this study, we use the Mean Absolute Percentage Error (MAPE) and the coefficient of determination ($R^2$) to evaluate the performance of our model. MAPE is a common metric used to assess the accuracy of ML models. It measures the average magnitude of absolute error as a percentage of the actual values, understanding how far predictions deviate from true values. The formula for MAPE is:

$$\text{MAPE} = \frac{1}{k} \sum_{i=1}^{k} \left| \frac{y_i - \hat{y}_i}{y_i} \right| \times 100 \tag{5}$$

Where $y_i$ is the actual value, $\hat{y}_i$ is the predicted value, and $k$ is the number of samples. MAPE expresses errors as a percentage, making it easy to interpret the relative size of errors.

$R^2$ indicates how well the model's predictions approximate the actual values. It measures the proportion of variance in the dependent variable predictable from the independent variables. $R^2$ values range from 0 to 1, where a value closer to 1 indicates a better fit. The formula for $R^2$ is:

$$R^2 = 1 - \frac{\sum_{i=1}^{k}(y_i - \hat{y}_i)^2}{\sum_{i=1}^{k}(y_i - \bar{y})^2} \tag{6}$$

where $\bar{y}$ is the mean of the actual values. $R^2$ evaluates how much of the variability in the actual data is captured by the model's predictions, with higher values indicating better performance.

## 4. Results and discussions

In this work, we first convert the MEA dump files into graphs, which is then fed into a GCNN-based ML model to predict the system's potential energy. Three case studies are conducted to evaluate the predictive accuracy of our model. All the data and codes for this work are provided in the Data Availability section.

### 4.1 Energy minimization profiles from hybrid MC/MD simulations

The results from our hybrid MC/MD simulations demonstrate significant changes in the atomic structure before and after the simulation. **Fig. 2(c)** illustrates these structural transformations. In the initial random alloy configuration, atoms are distributed randomly, with no apparent ordering. However, after thermal annealing at $T_a$= 350K, we observe the emergence of thermally stable structures, where the clustering of Ni-Ni and Co-Cr atoms is particularly favorable. This can be inferred from the highly negative values of their respective WC parameters, indicating a strong tendency for these atom pairs to cluster. As the annealing temperature increases, the atomic configuration exhibits greater disorder, as reflected in the diminished chemical ordering. For instance, at $T_a = 1150$K, we observe significantly less ordering, which is further supported by the WC parameter values showing a near-random distribution. **Fig. 2(a)** illustrates the pairwise WC parameters of the RSS system and systems after 60000 MC steps at different $T_a$. To capture the LCO of the alloy, we consider the 12 nearest neighbors of each atom to construct the edge list representation. This approach allows us to capture the local chemical ordering that is critical for understanding the material's properties. Dump files are extracted at every 200 MC steps, resulting in 300 distinct configurations for each annealing temperature. These configurations are then converted into graph-based representations, which form the input data for the GCNN-based ML model. The total potential energy of the system, normalized by the number of atoms, is used as the label for training the model. Similar procedures are followed for all annealing temperatures to generate the necessary data for the model.

### 4.2 Model's predictive performance for specific annealing temperatures

In the first case study, we evaluate the predictive performance of our model on data corresponding to individual annealing temperatures. Three models are trained using annealing temperatures of 450K, 650K, and 950K, as illustrated in **Fig. 3**. The variation in potential energy predicted by the model with the lowest MSE loss, compared to the actual data from hybrid MC/MD simulations, is shown in **Fig. 6(a)**. Additionally, **Fig. 8(a)** illustrates the model's predictive performance, with an $R^2$ value of 0.98 and a MAPE of 0.0210 for the 450K annealing temperature. For 650K, the $R^2$ value is 0.93 with a MAPE of 0.0499 and for 950K, the $R^2$ value is 0.96 with a MAPE of 0.0184. These results suggest that the model successfully captures the essential features of the training data and demonstrates strong predictive performance across varying annealing temperatures. Each model shows a good fit to its respective training data, confirming the robustness of the approach across different annealing temperature regimes.

## 4.3 Model's predictive performance for a set of different annealing temperatures

In the second case study, we explore the capability of a common model to distinguish between and accurately predict potential energies across different annealing temperatures when trained on a combined dataset of different annealing temperatures. Specifically, an ML model is trained using data from annealing temperatures of 450K, 650K, and 850K, as illustrated in **Fig. 4**. The variation in actual potential energy with MC steps, compared to the predictions generated by the model parameters with the lowest MSE loss model parameters during training is presented by **Fig. 6(b)**. Additionally, **Fig. 8(b)** illustrates the model's predictive performance across the different annealing temperatures. For the 450K annealing temperature, the model achieved an $R^2$ value of 0.97 with a MAPE of 0.0246, demonstrating excellent predictive accuracy. Similarly, for 650K, the $R^2$ value is 0.97 with a MAPE of 0.0216 and for 850K, the $R^2$ value is 0.96 with a MAPE of 0.0232. These consistently high $R^2$ values, close to 1, indicate that the model can explain almost all of the variance in the potential energy data across the three different annealing temperatures. The low MAPE values further confirm the model's precision. These results indicate that a common ML model, trained on a common dataset of multiple annealing temperatures, can accurately predict the potential energies across a range of temperature conditions.

## 4.4 Model's predictive performance for unseen annealing configurations

In the third case study, we examine the generalization capability of our model in predicting the properties of unseen structures, specifically investigating whether a single common model can predict the potential energies of structures not included in the training dataset. To this end, we divide our hybrid MC/MD simulation data into two groups: a training dataset and a testing dataset. The training dataset comprises data from annealing temperatures of 350K, 650K, 850K, and 1150K, while the testing dataset consists of data from other annealing temperatures. The model is trained on the training. These model parameters are then used to predict the potential energies of the structures of the train and test set, as illustrated in **Fig. 5**. The variation in actual potential energy with MC steps, compared to the predictions from the model parameters yielding the lowest MSE loss during training, is presented in **Fig. 7** for both training and testing datasets. Additionally, **Fig. 9** illustrates the model's predictive performance across the different annealing temperatures. We observe that the $R^2$ value drops as the annealing temperature increases for both the training and testing sets. The model demonstrates strong predictive performance for the lower annealing temperatures in both training and testing sets, accurately capturing the potential energy variation. However, as the annealing temperature increases, the system's potential energy decreases sharply and reaches a plateau more quickly. This poses a challenge for the model because the rapid drop in potential energy at higher temperatures results in fewer data points at higher energy levels. Consequently, the model receives less information about these higher energy states, making it more difficult to predict potential energy accurately during the initial stages of the MC steps at elevated temperatures. Although the model effectively captures the overall trend of potential energy variation, it struggles to predict the higher energy levels due to the limited amount of data available at these high annealing temperatures. This decline in predictive accuracy at higher temperatures suggests that while the model generalizes well to lower annealing temperature conditions, it faces limitations in accurately predicting the high-energy data at higher annealing temperatures.

## 4. Conclusion

In this study, the potential energy of an MEA in hybrid MC/MD simulations at varying annealing temperatures is predicted using a GCNN-based ML model. The hybrid MC/MD simulations generate dump files that are converted into graphs. The node features are designated, which include atom type and velocity. The model's efficacy is assessed through the implementation of three case studies. The model exhibits satisfactory performance on the training data in the initial case study, which involves the training of three distinct models using data from three different annealing temperatures. The model's performance on a combined annealing temperature dataset is evaluated in the second case study by training it on data from all three annealing temperatures. The results of the model are satisfactory, particularly in terms of predicting the variation of potential energy regarding MC phases.

Lastly, in the third case study, the model is trained on some annealing temperatures and its performance is evaluated on the testing set, where it exhibited satisfactory performance on both the training and testing data. It is essential to illustrate in this investigation that the graph representations of MEA can be utilized for predicting properties. Our model accurately predicts the variation in potential energy that is correlated with the system's LCO. We accomplish satisfactory performance in estimating the potential energy by establishing edges between an atom and its 12 nearest neighbors. This research offers valuable insights into the graph modeling of MEAs and HEAs, demonstrating the effective capture of LCO variation by our model. In addition, this research establishes the foundation for the development of more sophisticated modeling methods for HEAs and MEAs that are based on graphs.

## Author contribution statement

**Mashaekh Tausif Ehsan:** Conceptualization, Methodology, Investigation, Validation, Data curation, Software, Formal analysis, Writing – original draft, Writing – review & editing. **Saifuddin Zafar:** Investigation, Visualization, Writing – review & editing. **Apurba Sarker:** Methodology, Software. **Sourav Das Suvro:** Writing – review & editing. **Mohammad Nasim Hasan:** Project administration, Resources, Supervision, Writing - review & editing.

## Data availability

All the data and codes for this work are available on GitHub: https://github.com/mashaekh-tausif/MCMD-GCNN

## Acknowledgement

This research is based in part on the methodology developed by Noda et al. [51], licensed under Creative Commons Attribution 4.0 International (CC BY 4.0) (https://creativecommons.org/licenses/by/4.0/). This work has been modified from the original to suit the unique requirements of our study. Unlike the original work, we performed hybrid MC/MD simulations of MEA. Additionally, we used the 12 nearest neighbors to construct the edge list for capturing the LCO, which differs from the original method. Additionally, we used atom type and atom absolute velocity as node features. We converted the node feature matrix and edge list into a PyTorch Geometric Data object and used DataLoader for effective batching and shuffling, which was not used in the previous study. Instead of the SAGEConv used by Noda et al., we utilized GCNConv, and our architecture differs in terms of the number of layers, epochs, and learning rate. We also did not use a learning rate scheduler in our model. Finally, we evaluated the performance of our model using R2, MAPE, and parity plots, which were not employed in the previous work.

## List of Figure Captions

| | |
|---|---|
| **Fig. 1** | Diagram of the machine learning model architecture built on Graph Convolutional Networks used in this study. |
| **Fig. 2** | (a) Pairwise WC parameter at different annealing temperatures. (b) Schematic diagram of Hybrid MC/MD method (c) Representative configurations of the RSS system, along with configurations of systems annealed at 350K and 1350K, respectively. |
| **Fig. 3** | Schematic diagram of graph construction, training steps, and predictions in the first case study. |
| **Fig. 4** | Schematic diagram of graph construction, training steps, and predictions in the second case study. |
| **Fig. 5** | Schematic diagram of graph construction, training steps, and predictions in the third case study. |
| **Fig. 6** | Variation of potential energy with MC steps showing both actual and predicted values in (a) the first case study (b) the second case study |
| **Fig. 7** | Variation of potential energy with MC steps, showing both actual and predicted values in the third case study in (a) training data, (b) testing data. |
| **Fig. 8** | Parity plots along with $R^2$ and MAPE values for (a) the first case study, (b) the second case study. |
| **Fig. 9** | Parity plots along with $R^2$ and MAPE values for the third case study predicting (a) training data (b) testing data . |

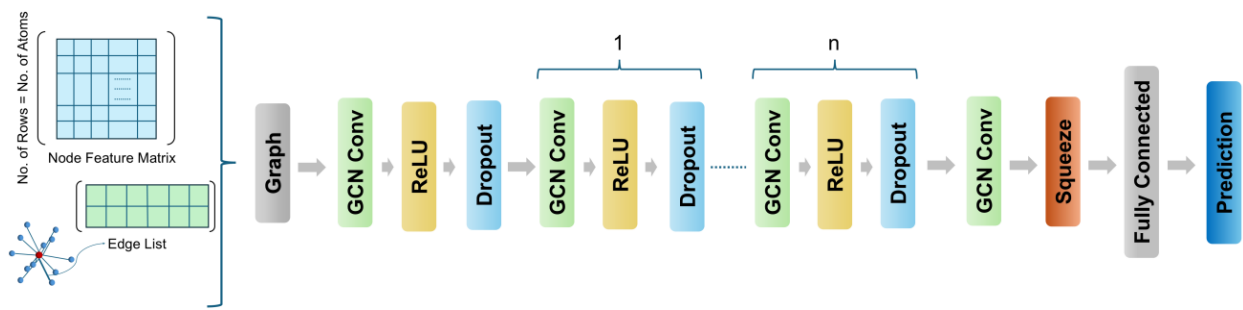

**Fig. 1.** Diagram of the machine learning model architecture built on Graph Convolutional Neural Networks used in this study.

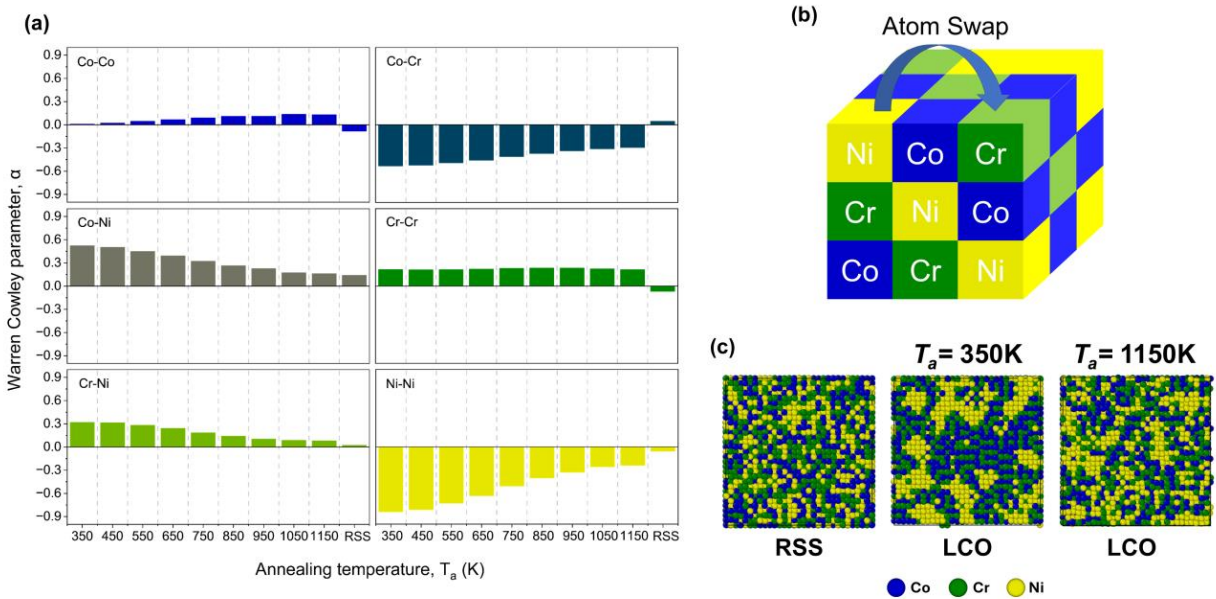

**Fig. 2.** (a) Pairwise WC parameter at different annealing temperatures. (b) Schematic diagram of Hybrid MC/MD method (c) Representative configurations of the RSS system, along with configurations of systems annealed at 350K and 1350K, respectively.

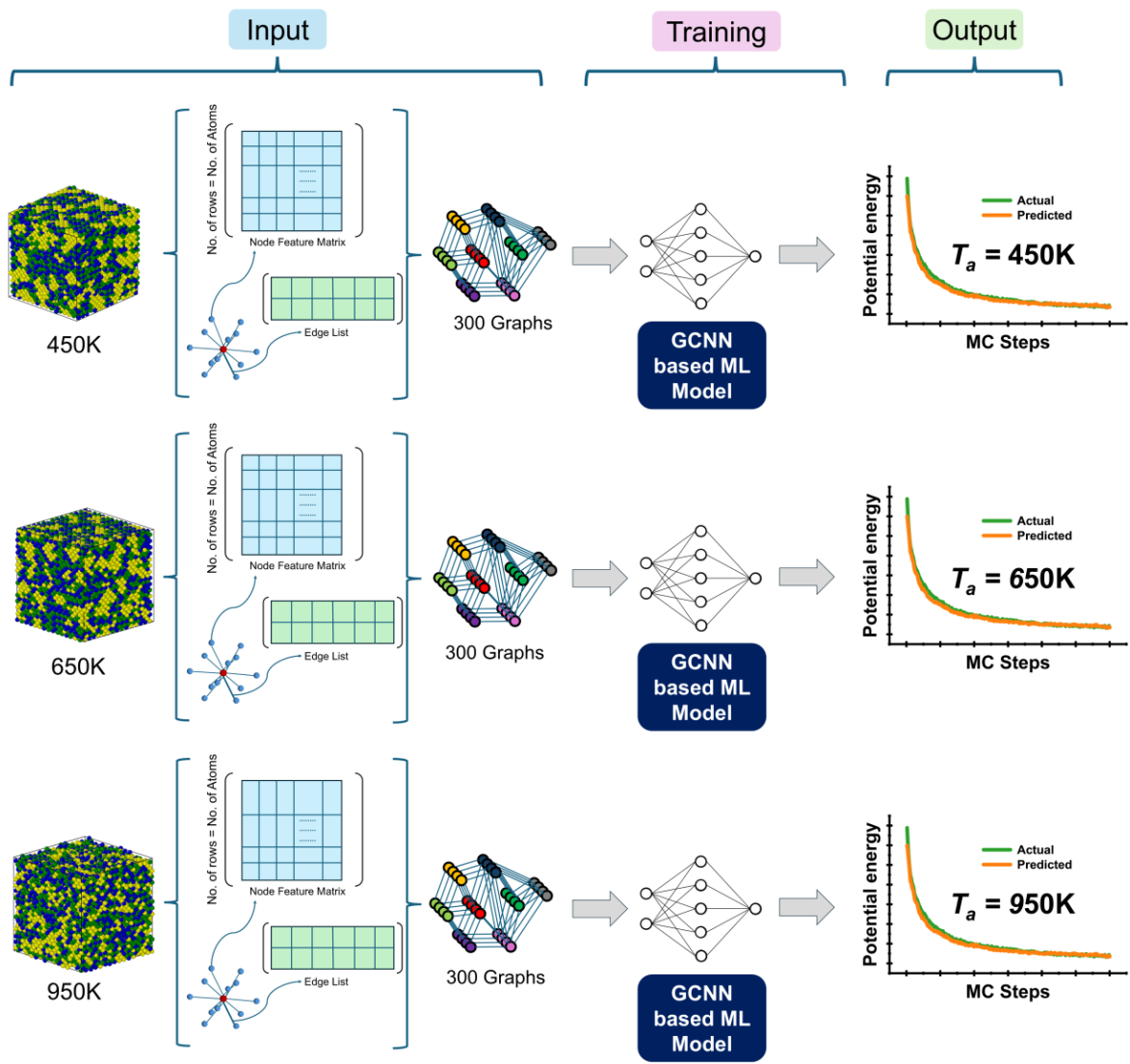

**Fig. 3.** Schematic diagram of graph construction, training steps, and predictions in the first case study

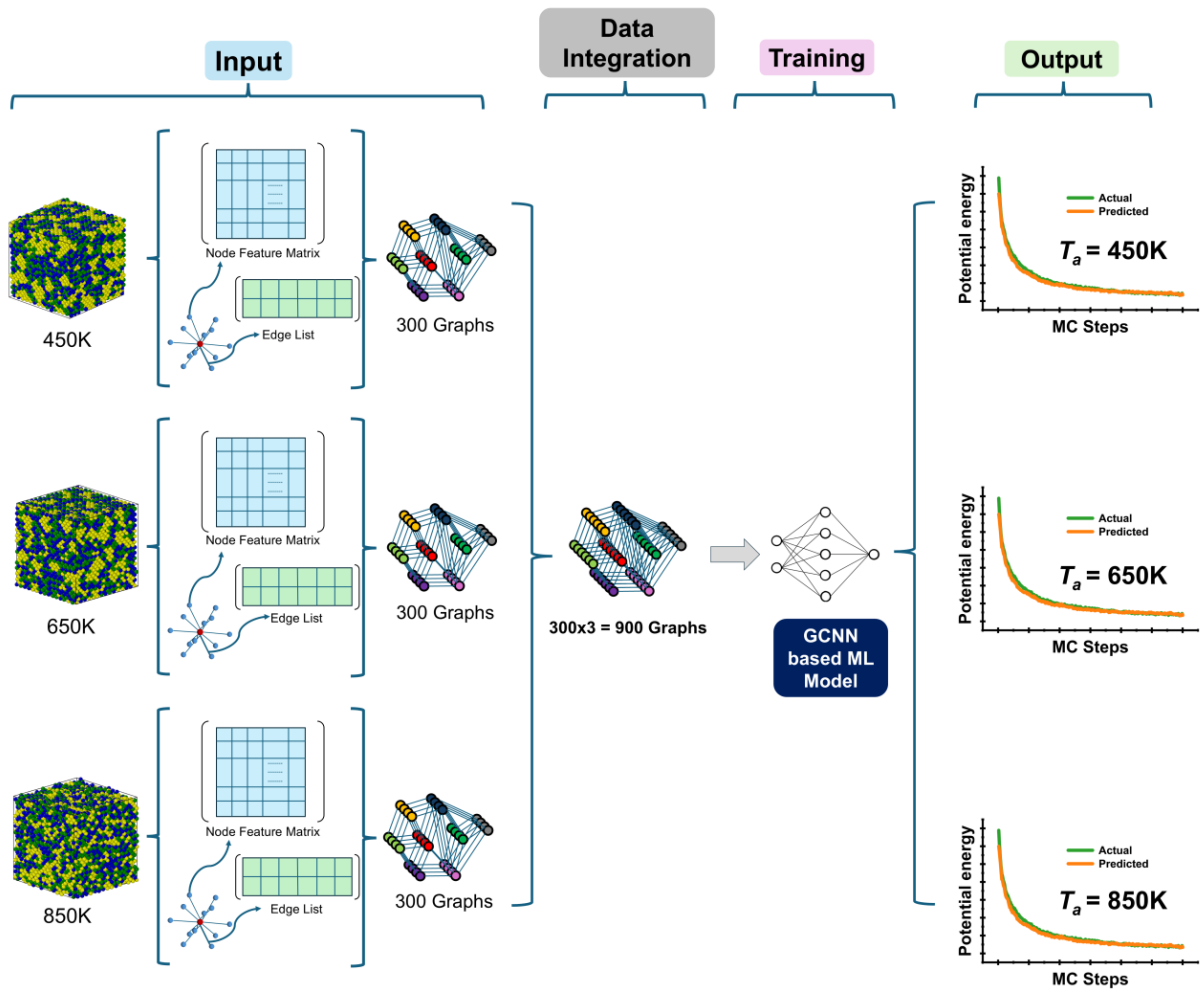

**Fig. 4.** Schematic diagram of graph construction, training steps, and predictions in the second case study.

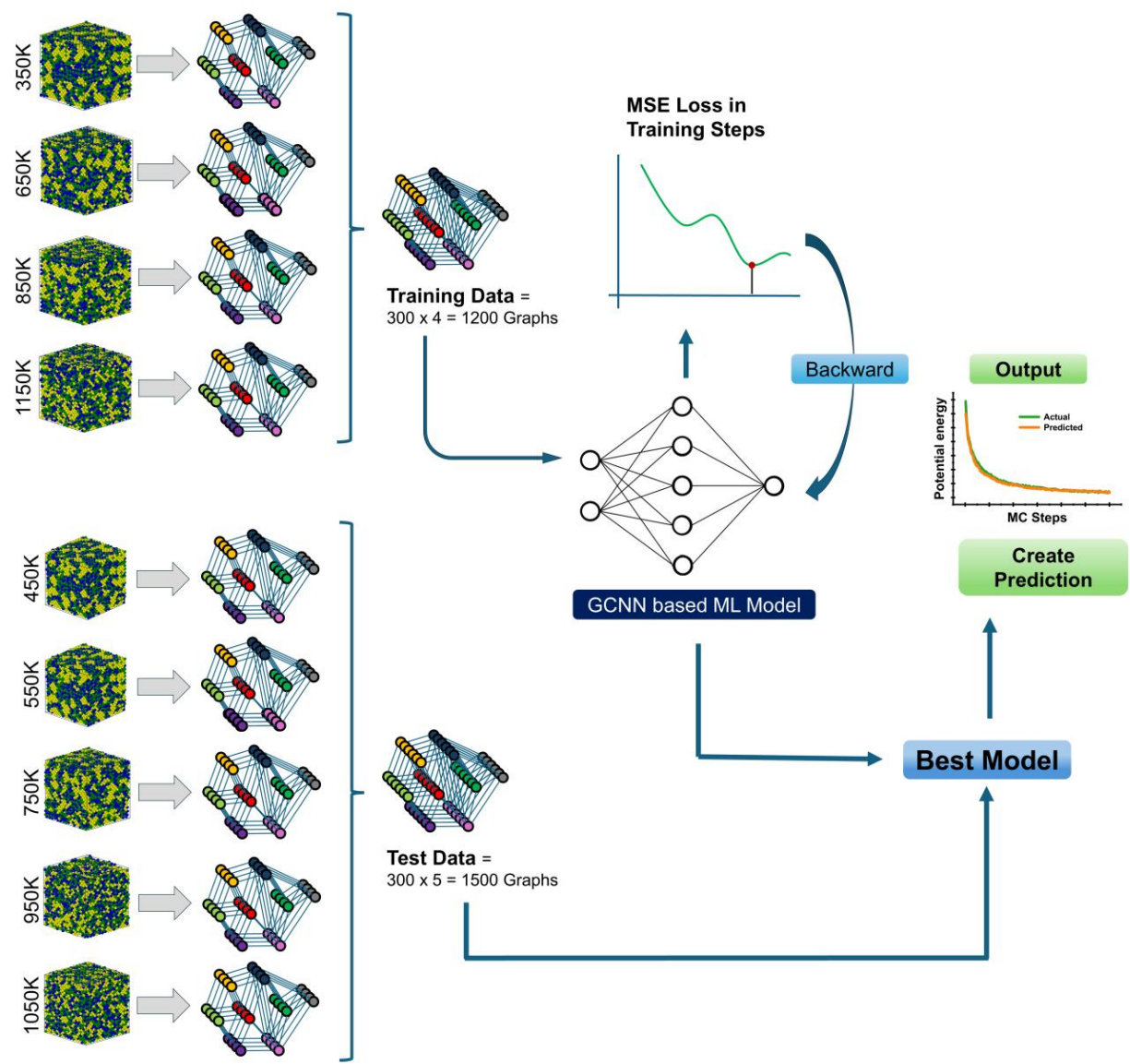

**Fig. 5.** Schematic diagram of graph construction, training steps, and predictions in the third case study

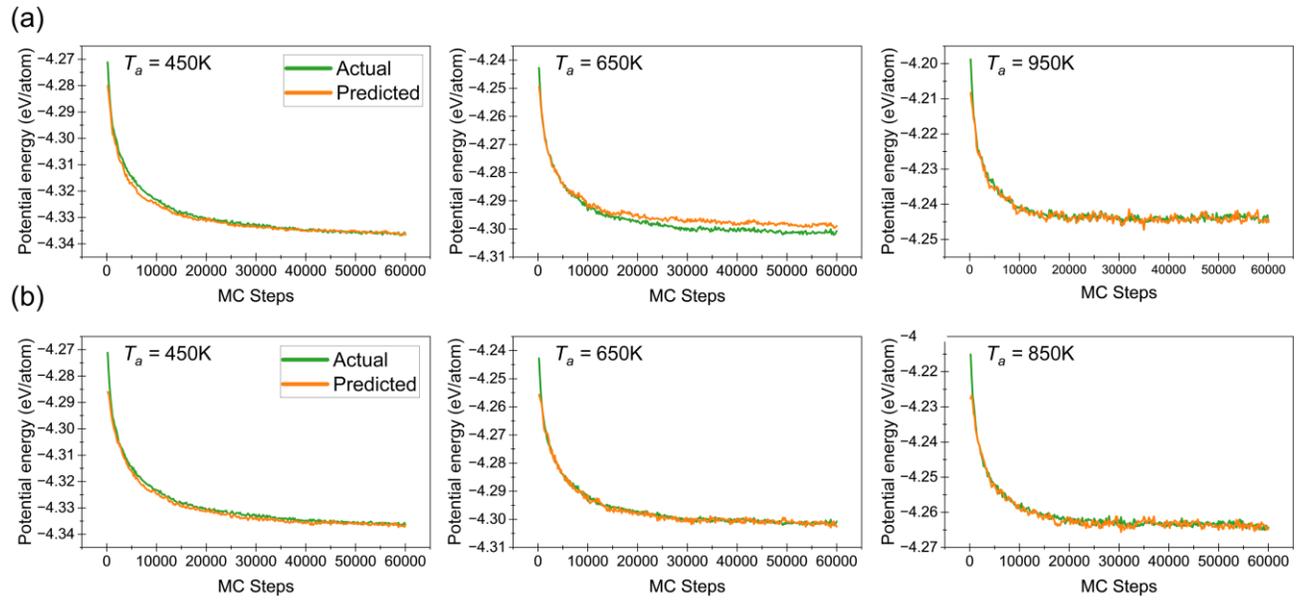

**Fig. 6.** Variation of potential energy with MC steps showing both actual and predicted values in (a) the first case study (b) the second case study

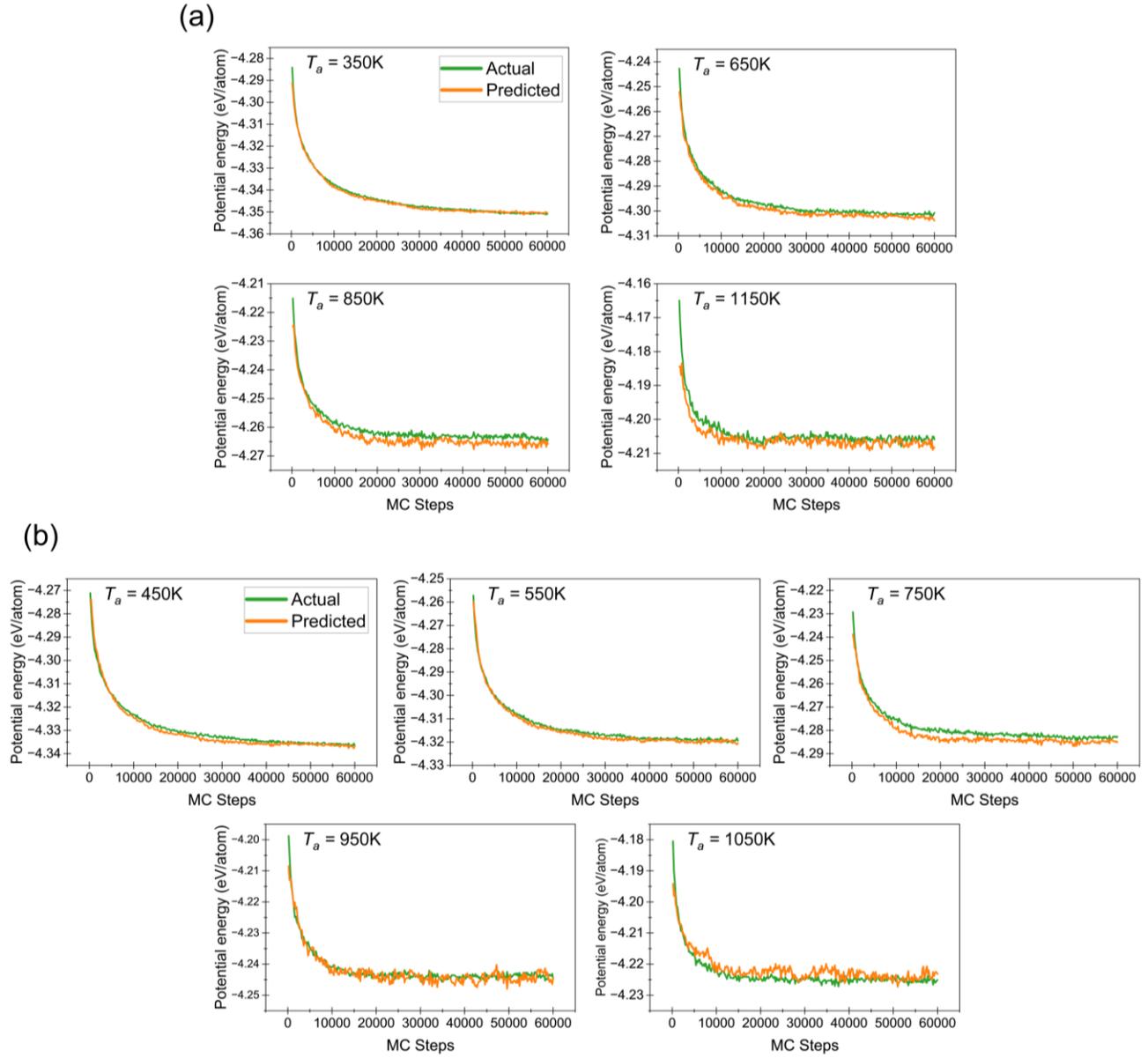

**Fig. 7.** Variation of potential energy with MC steps, showing both actual and predicted values in the third case study (a) training data, (b) testing data

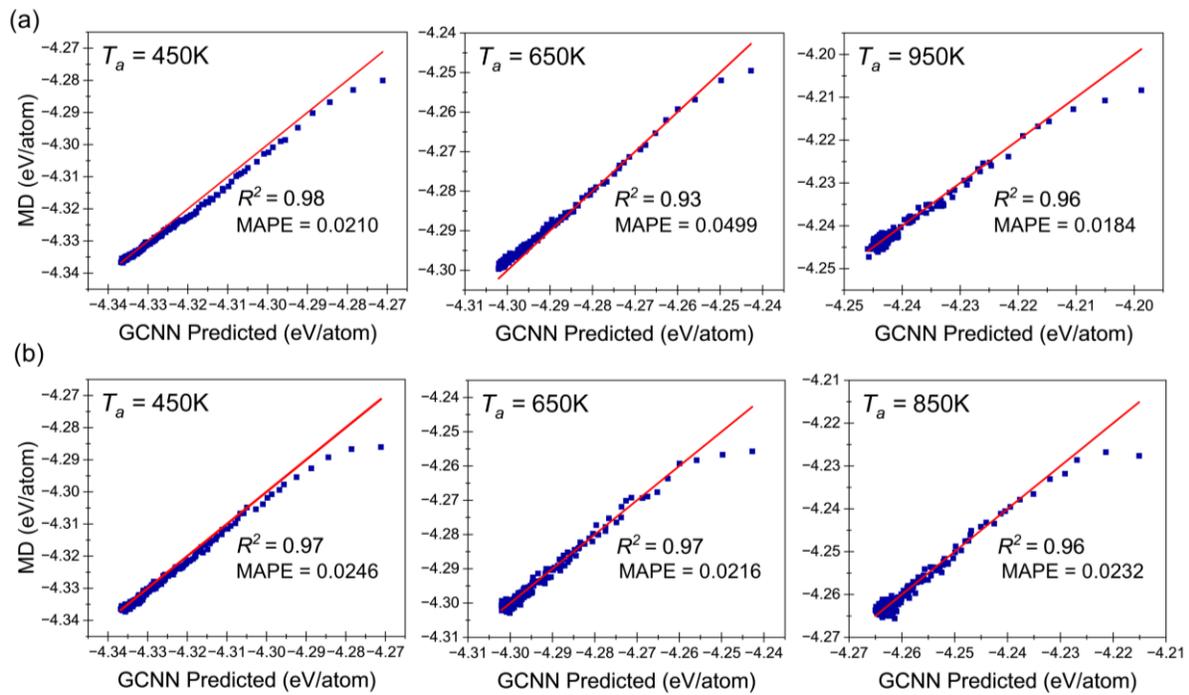

**Fig. 8.** Parity plots along with $R^2$ and MAPE values for (a) the first case study, (b) the second case study

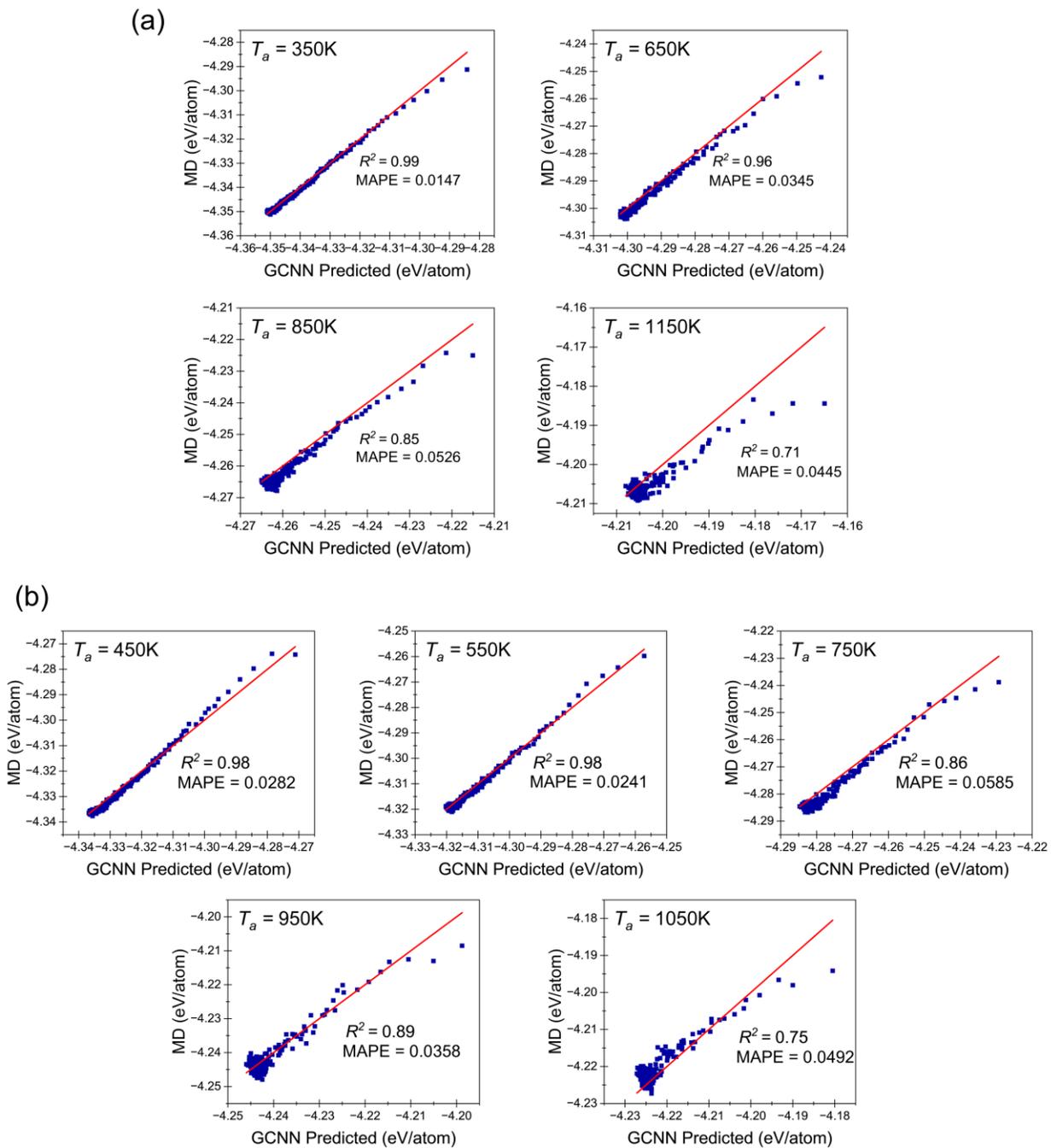

**Fig. 9.** Parity plots along with $R^2$ and MAPE values for the third case study predicting (a) training data (b) testing data